\documentclass{aa}

\usepackage{natbib}
\usepackage{longtable}
\usepackage{lscape}
\usepackage{amsmath}
\usepackage{txfonts}
\usepackage{color}
\usepackage{url}
\usepackage{xspace}

\usepackage{graphicx}
\bibpunct{(}{)}{;}{a}{}{,}

\newcommand{\kev}{keV\xspace}
\newcommand{\xmm}{\textsl{XMM-Newton}\xspace}
\newcommand{\rosat}{\textsl{ROSAT}\xspace}
\newcommand{\chandra}{\textsl{Chandra}\xspace}

\begin{document}
   \title{A new luminous supersoft X-ray source in NGC~300}

   \author{S. Carpano
          \inst{1}
          \and
          J. Wilms
          \inst{2}
	  \and
	  M. Schirmer
	  \inst{3}
	  \and
          E. Kendziorra
          \inst{1}}

   \offprints{S. Carpano, e-mail: scarpano@sciops.esa.int}
   \institute{Institut f\"ur Astronomie und Astrophysik, Abteilung
   Astronomie,  Universit\"at T\"ubingen, Sand 1, 72076 T\"ubingen, Germany   
	      \and
   Department of Physics, University of Warwick, Coventry, CV4 7AL,
   United Kingdom
             \and
   Isaac Newton Group of Telescopes, 38700 Santa Cruz de La Palma, Spain
   }
   
   \date{Received 15 December 2005/ Accepted 27 July 2006}
   
   \abstract{\emph{Context}. We report the discovery of a new luminous supersoft
     source, \object{XMMU J005455.0$-$374117}, in the nearby spiral
     galaxy NGC~300, in \xmm observations performed on 2005 May 22 and 
     on 2005 November 25.
     The source is not present in \rosat data nor in the previous \xmm
     observations of 2000 December/2001 January.  The unique luminous supersoft source, \object{XMMU
       J005510.7$-$373855}, detected in the 1992 May/June \rosat data and 
       in the 2000/2001
     \xmm data, fell below detectability.  This source already
     appeared highly variable in  \rosat observations. \emph{Aims}. We report
     on the temporal and spectral analysis of this new supersoft source
     (SSS) and compare its properties with the previous known SSS.
     \emph{Methods}. We present the  light curves of the SSS, model its
     spectrum and estimate the corresponding flux and
     luminosities. \emph{Results}. The light curve of XMMU J005455.0$-$374117  does
     not show large fluctuations  in any of the observations and its
     spectrum  can be  modelled with an absorbed blackbody with
     $kT\sim60$\,eV. The corresponding bolometric luminosity is 
     $8.1^{+1.4}_{-4.5}\times10^{38}\,\text{erg}\,\text{s}^{-1}$ in the first
     observation and drops to
     $2.2^{+0.5}_{-1.4}\times10^{38}\,\text{erg}\,\text{s}^{-1}$ six months
     later. No optical source  brighter than
     $m_\text{V}\sim21.7\,\text{mag}$ is found coincident with its
     position. \emph{Conclusions}. The luminosity of 
     these two SSSs is higher than what has been found for `classical'
     SSSs. Their nature could be explained by beamed emission from
     steady nuclear burning of hydrogen onto white dwarfs, or
     accretion onto stellar-mass black hole with matter outflow or
     observed at high inclination angle. The presence of an
     intermediate-mass black hole seems unlikely in our case.

      }

\maketitle
%
%____________________________________________________________________________

\section{Introduction}
\label{sec:int}

Supersoft sources (SSSs) were first discovered in the Large Magellanic
Cloud with the \textsl{Einstein} Observatory \citep{Long1981}, 
including the two prototypes \object{CAL83} and \object{CAL87}.  SSSs are
characterised by very soft thermal spectra, with temperatures
typically below 100\,eV, and have bolometric luminosities in excess of
$10^{36}\,\text{erg}\,\text{s}^{-1}$.  These sources can be divided
 into two groups. The first one includes the ``classical'' SSSs,
which are characterized by bolometric luminosities in the range
$10^{36}$--$2\times10^{38}\,\text{erg}\,\text{s}^{-1}$.  The most
promising model to explain the flux emitted from such SSSs was
proposed by \cite{vandenHeuvel1992} to be a steady nuclear burning
of hydrogen accreted onto white dwarfs with masses in the range of
0.7--1.2\,M$_{\odot}$. These sources are fairly common: 57 sources
have been catalogued by
J.~Greiner\footnote{\url{http://www.aip.de/~jcg/sss/ssscat.html}}
up to 1999 December, but since then many more were discovered in
distant galaxies by \xmm and \chandra (see below). The second
group includes SSSs for which the luminosity exceeds the Eddington
limit for a $1.4\,\text{M}_\odot$ compact object, therefore excluding
unbeamed emission from steady nuclear burning of hydrogen accreted
onto a white dwarf. These sources are much less common.  For
ultraluminous SSSs, i.e., SSS with bolometric luminosities exceeding
$10^{39}\,\text{erg}\,\text{s}^{-1}$, models involving
intermediate-mass black holes \citep{Kong2003, Swartz2002, Kong2004}
or stellar-mass black holes with matter outflow \citep{Mukai2005} have
been invoked.

According to \cite{Greiner2004}, 25 SSSs have been discovered in
\object{M31} of which a large fraction (30\%) are found to be
transient sources, with turnoff and turnon times of the order of a few
months.  Their luminosities are in the range
$10^{36}$--$10^{38}\,\text{erg}\,\text{s}^{-1}$.  \cite{DiStefano2003}
found 16 SSSs in \object{M101}, 2--3 in \object{M51}, 10 in
\object{M83} and 3 in \object{NGC~4697}. Of these sources, 11 have
bolometric luminosities $>10^{38}\,\text{erg}\,\text{s}^{-1}$, from
which six are brighter than $10^{39}\,\text{erg}\,\text{s}^{-1}$. In
\object{M81}, \cite{Swartz2002} found 9 SSSs, including two with
bolometric luminosities $>10^{38}\,\text{erg}\,\text{s}^{-1}$ and one
$>10^{39}\,\text{erg}\,\text{s}^{-1}$.  According to
\cite{DiStefano2003}, normal SSSs in spiral galaxies appear to be
associated with the spiral arms. The most luminous SSSs, however, have
been found either in the arms, bulge, or disk \citep{Swartz2002}, as
well as in halos \citep{DiStefano2003b}. SSSs have been found either
in spiral (e.g. M31, M101, M83,
M81, \object{M104}, \& NGC~300), elliptical (e.g.,
NGC~4697), interacting (e.g., M51 and
\object{NGC~4038}/\object{NGC~4039}, i.e., the Antennae) or irregular
galaxies (e.g., \object{LMC} and \object{SMC}).

In this paper we report  the discovery of a luminous
($>2\times10^{38}\,\text{erg}\,\text{s}^{-1}$) SSS,
XMMU~J005455.0$-$374117, in the spiral galaxy NGC~300.  This galaxy is
a normal dwarf galaxy of type SA(s)d located at a distance of
$\sim$1.88\,Mpc \citep{Gieren2005}. The galaxy is seen almost face-on
and has a low Galactic column density
\citep[$N_\text{H}=3.6\times10^{20}\,\text{cm}^{-2}$;][]{Dickey1990}.
The major axes of its $D_{25}$ optical disk are 13.3\,kpc and 9.4\,kpc
\citep[$22'\times 15'$;][]{deVaucouleurs1991}.  NGC~300 was
observed by \rosat between 1991 and 1997 for a total of 46\,ksec with
the Position Sensitive Proportional Counter and 40\,ksec with the High
Resolution Imager.  One SSS, XMMU J005510.7$-$373855, was present in
these observations \citep{Read1997}. This source was visible in 1992
May and June but not in 1991 December \citep{Read2001}.  The spectrum
was well described with a thermal bremsstrahlung model with $kT \sim
0.1$\,keV \citep{Read1997}.

More recently, we observed NGC~300 with \xmm during its orbit 192
(2000 December 26; 37\,ksec on source time) and orbit 195 (2001
January 1; 47\,ksec on source time).  The results of these observations
have been presented by \cite{Carpano2005}.  A deep analysis of the SSS
XMMU J005510.7$-$373855 as seen in these observations was also
performed by \cite{Kong2003}.  These authors report that during the
6\,days between the two \xmm pointings the source went from a ``high
state'' to a ``low state'', and that a 5.4\,h periodicity was found
during the low state. More information about this source will be given 
in Sect.~\ref{sec:discus}. Recently, \xmm re-observed NGC~300 on
2005~May~22 (orbit 998) and on 2005~November~25 (orbit 1092), for
35\,ksec each.

In this paper, we focus on the analysis of a new SSS, which was
present in the 2005 \xmm observations, and compare its properties with
the previously known SSS.  For simplicity, we will refer to 
XMMU~J005510.7$-$373855 as SSS$_{1}$ and XMMU J005455.0$-$374117 as
SSS$_{2}$ in this work.  Sect.~\ref{sec:obs} describes
the observations and the reduction of the \xmm data. In
Sect.~\ref{sec:res} we present the spectral and timing analysis of
SSS$_2$. We discuss  the nature of these SSSs in
Sect.~\ref{sec:discus} and conclusions are given in
Sect.~\ref{sec:conc}.

%_____________________________________________________________________________

\section{Observations and data reduction}
\label{sec:obs}
For the 2005 May and November \xmm observations, the EPIC-MOS
\citep{Turner} and EPIC-pn \citep{Strueder} cameras were operated in
their full frame mode with the medium filter.  The 
EPIC-pn camera was centered on the previously known SSS$_{1}$
($\alpha_\text{J2000.0}=00^\text{h}55^\text{m} 10\fs{}7$ and
$\delta_\text{J2000.0}=-37^\circ 38' 55\farcs 0$).  The data reduction
was identical to that used in our analysis of the previous \xmm
observations \citep{Carpano2005}, except that version 6.5.0 of the
\xmm Science Analysis System (SAS) and recent calibration files were
 used.  After screening the MOS data for proton flares using the
standard procedures described by the \xmm team, a total of 30\,ksec of
low background data remained for  revolution 998.  The same
good time intervals were then also used for the EPIC-pn data, leaving
26\,ksec of low background.  No high background was present in the
data of orbit 1092 where the exposure time was of 36\,ksec for the MOS
and 31\,ksec for the pn data.

Using the SAS \texttt{edetect\_chain} task, which performs a maximum
likelihood source detection, SSS$_2$ is detected with a likelihood of
$5.2\times 10^3$ in the combined observations of orbits 998 and 1092.
Following \citet{Carpano2005}, we improve the X-ray positions by cross
correlating positions between X-ray sources and their optical
counterparts. The revised coordinates of the source are
$\alpha_\text{J2000.0}=00^\text{h}54^\text{m} 55\fs{}0$ and
$\delta_\text{J2000.0}=-37^\circ 41' 17\farcs 0$, with an uncertainty
of $0\farcs 64$.

%_____________________________________________________________________________

\section{Timing and spectral analysis of SSS$_2$}
\label{sec:res}
\begin{figure}
\resizebox{\hsize}{!}{\includegraphics{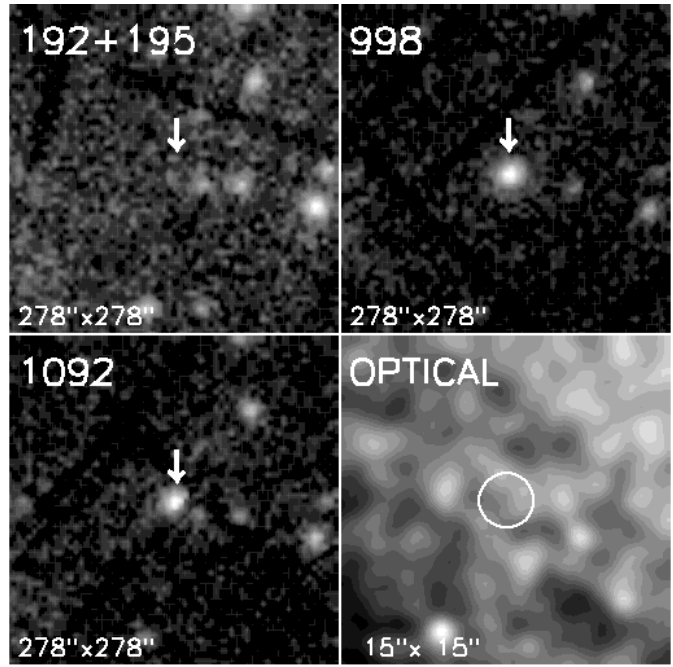}}
\caption{$278''\times278''$ images of SSS$_{2}$, in the 0.2--2.0 \kev 
  band from the different \xmm observations (revolution 192+195, 998,
  and 1092) and $15''\times15''$ optical image centered on the X-ray
  position of the source (the circle represents the $2\sigma$
  uncertainty of the X-ray position).}
 \label{fig:image}
\end{figure}

Fig.~\ref{fig:image} shows the combined MOS/pn images centered on
SSS$_2$ taken during \xmm observations (revolution 192 and
195, 998, and 1092). SSS$_{2}$ is located close to the center of the
galaxy, at a projected distance of $\sim$0.24\,kpc.  We also show the
combined BVR optical image of the field taken with the 2.2\,m MPG/ESO
telescope on La Silla.  See \citet{Carpano2005} and
\citet{Schirmer2003} for a description of the optical data and their
reduction.

SSS$_{2}$ was clearly visible in revolutions 998 and 1092 but not in
revolutions 192 and 195, where the detection limiting luminosity is
$\sim1.3\pm0.6\times10^{36}\,\text{erg}\,\text{s}^{-1}$ and
$\sim1.1\pm0.5\times10^{36}\,\text{erg}\,\text{s}^{-1}$, respectively
(assuming a blackbody model with $kT\sim60$\,eV and
$N_\text{H}=10^{21}\,\text{cm}^{-2}$, with a $4\sigma$ confidence
level).  On the other hand, SSS$_{1}$ was detected in revolution 192
and 195 \citep{Kong2003} but not in the last two revolutions where the
detection limiting flux is of
$\sim1.2\pm0.6\times10^{36}\,\text{erg}\,\text{s}^{-1}$ in both cases.
SSS$_{2}$ has not been detected in any of the 5 \rosat observations,
although it would have been detectable in the first four \rosat
observations (where the detection limit was
$<3.3\pm1.1\times10^{37}\,\text{erg}\,\text{s}^{-1}$) if it had had a
luminosity similar to what has been found in the \xmm data.  In the
optical images, including data from the Optical Monitor on \xmm, no
counterpart has been found to be bright enough to coincide with either
of the SSSs and nor does the SIMBAD catalogue list possible
counterparts.  Because SSS$_{2}$ is located close to the center of
NGC~300, the optical detection limit is high: 21.7\,mag, 21.7\,mag,
21.4\,mag in the B, V, and R band respectively.  For SSS$_{1}$,
located in one of the spiral arms of the galaxy, no optical
counterpart brighter than $m_\text{V}=24.5\,\text{mag}$ coincides with
the source, therefore excluding  the presence of an O or early B
companion star \citep[see][ for the optical field around the
source]{Carpano2005}.

\begin{figure}
  \resizebox{\hsize}{!}{\includegraphics[bb=20 8 333 297,clip=true]{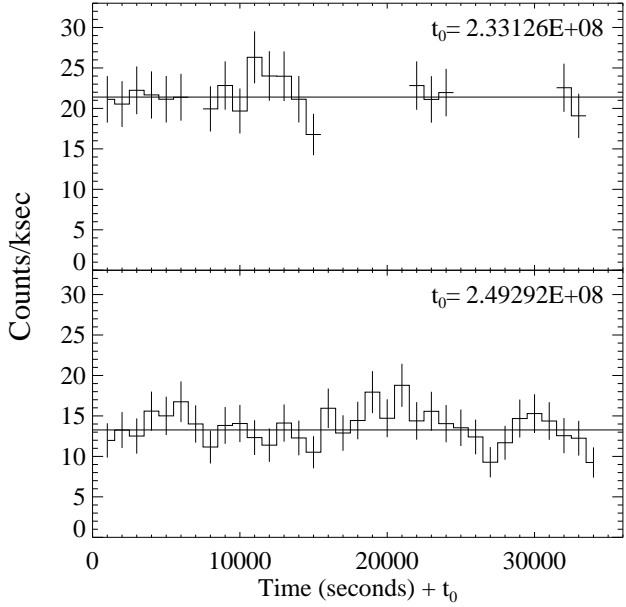}}
 \caption{MOS and pn 0.2--2.0\,keV light curve of SSS$_{2}$ in
   revolution 998 (top) and 1092 (bottom).  Periods of high background
   have been excluded from the data. The horizontal line shows the
   fitted mean value.  Times given are barycentric and measured in
   seconds from 1998 January 1 (MJD 50814.0).}
 \label{fig:light}
\end{figure}

Fig.~\ref{fig:light} shows the light curve of SSS$_{2}$ in revolutions
998 and 1092.  Periods of high background have been excluded from the
data. The light curve does not present large fluctuations. To test the
significance of the source variability, we fit a constant value to the
light curves (binned to 1000\,s), and, from the resulting $\chi^2$ (9
for 18\,dof, in rev. 998, and 31 for 33\,dof, in rev. 1092), we find
that the source is variable with a probability of 5\% and 42\%, for
revolution 998 and 1092, respectively.  Using periodograms and
epoch-folding, no short-time periodic signal was found on timescales
from 5\,sec to 30\,ksec.

\begin{figure}
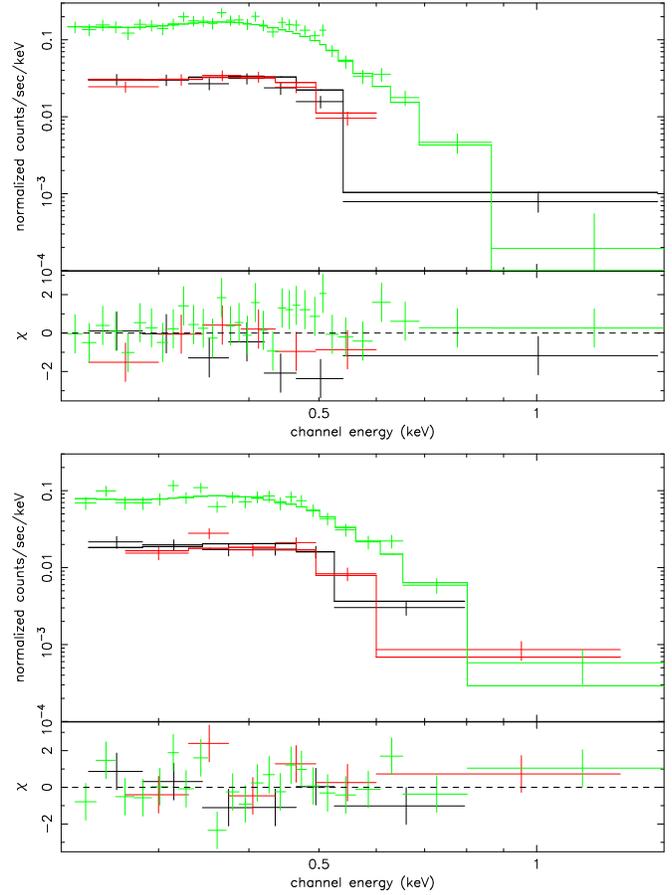

  \resizebox{\hsize}{!}{\includegraphics[bb=113 44 563 708,clip=true,angle=-90]{5609fig3.ps}}
  \resizebox{\hsize}{!}{\includegraphics[bb=113 44 563 708,clip=true,angle=-90]{5609fig4.ps}}
 \caption{pn and MOS spectra of source SSS$_{2}$ observed in
   revolution 998 (top) and 1092 (bottom), and the best fit spectral
   model. Bottom of each spectrum: residuals expressed in $\sigma$.}
 \label{fig:spec}
\end{figure}

\begin{table*}
 \centering
 \caption{Results of the spectral fits for SSS$_{2}$,
   using an absorbed blackbody model (\texttt{phabs+bbody}, in XSPEC),
   where $N_\text{H}$ is the column density of neutral hydrogen, $kT$
   the temperature and $\chi^2_{\nu}$/dof is the reduced chi-square
   and the number of degrees of freedom. The corresponding
   0.2--2\,\kev flux, and absorbed luminosities, as well as the
   bolometric luminosities, are shown in the last three
   columns. Uncertainties are given at a 90\% confidence level, except
   for the bolometric luminosity, where they are at 99\%.}   
  \label{tab:spec_fit}
 \begin{tabular}{lllllll}
 \hline
source (rev.) &  $N_\text{H}(\times 10^{20}\,\text{cm}^{-2})$  & $kT$ (eV) & $\chi^2_{\nu}$/dof &  $F_{0.2-2}$ (\,cgs) &  $L_{0.2-2}^{\text{obs}}$ (\,cgs) &   $L^{\text{bol}}$ (\,cgs)\\[3pt]
\hline
SSS$_{2}$ (998) & $8.02^{+2.20}_{-1.53}$ & $54^{+3}_{-4}$ & $1.03$/$43$ & $1.18^{+1.07}_{-0.96}\times 10^{-13}$ & $4.99^{+4.54}_{-4.06}\times 10^{37}$ &  $8.12^{+1.39}_{-4.47}\times 10^{38}$\\[3pt]
SSS$_{2}$ (1092)& $5.82^{+1.97}_{-2.39}$ & $62^{+6}_{-4}$ & $1.16$/$31$ & $0.71^{+0.39}_{-0.49}\times 10^{-13}$ & $3.01^{+1.66}_{-2.07}\times 10^{37}$ &  $2.21^{+0.45}_{-1.40}\times 10^{38}$\\[3pt]
  \hline
   \end{tabular}
\end{table*}

Fig.~\ref{fig:spec} shows the spectra and the best fit spectral model
of SSS$_{2}$ for revolutions 998 and 1092. The data were
  binned to have at least 35 counts in each energy bin. Results of the
spectral fits, the corresponding 0.2--2\,\kev flux absorbed
luminosities, and the bolometric luminosities are given in
Table~\ref{tab:spec_fit}.  During revolution 1092 SSS$_{2}$ is
situated on an EPIC-pn CCD gap.  For this observation the
flux/luminosities are calculated from the MOS1 data alone.  We tried
to describe the data with several spectral models, including
bremsstrahlung, power-law, blackbody and disk blackbody. The blackbody
and disk blackbody are the only models that provide a reasonable value
for the reduced $\chi^2$, ($\chi^2_{\nu}<1.2$). The spectral fit
parameters and flux resulting from the disk blackbody model are very
similar to that provided by the simple blackbody model.  Therefore,
for simplicity, we assume the simple blackbody model in the rest of
the paper.  As suggested by \cite{Mukai2005}, we also tried an ionized
model for the absorption (implemented in XPSEC as \texttt{absori}),
but the spectral parameters corresponding to this model cannot be
constrained and the best-fit model is a neutral absorber (ionization
state $\xi=0$).

From the results of Table~\ref{tab:spec_fit} we see that within
6~months the absorbing column slightly decreased (although the
associated errors are very large), the temperature increased, and the
observed luminosity dropped by a factor of $\sim$1.7.

Bolometric luminosities of supersoft sources are difficult to
determine due to the large uncertainty associated with the absorbing
column.  In our case, a blackbody model associated with photo-electric
absorption provides a low $\chi^2$ value ($\chi^2_\nu<1.05$) and any
other suitable model does not significantly change the bolometric
luminosity.  Using these considerations, we therefore conclude that
the high luminosity of SSS$_{2}$, which is above the Eddington
luminosity of a white dwarf, excludes the presence of unbeamed
emission from steady nuclear burning of hydrogen accreted onto a white
dwarf. In the next section we discuss the models that could
explain the nature of the SSSs observed in NGC~300 after summarizing 
the information we have on the previously known SSS$_{1}$.

%_____________________________________________________________________________

\section{The nature of SSS$_{1}$ and SSS$_{2}$}
\label{sec:discus}
\subsection{What do we know about SSS$_{1}$?}
SSS$_{1}$ is a transient source: the observed luminosity changed by at
least a factor of $\sim$35 (we measured an observed luminosity of
$4\times10^{37}\,\text{erg}\,\text{s}^{-1}$ while the \xmm detection
threshold is $\sim 1.2\times10^{36}\,\text{erg}\,\text{s}^{-1}$).  The
source has been detected in the \rosat data of 1992 May/June
\citep{Read2001} and in the \xmm data of 2002 December/2001 January
\citep{Kong2003,Carpano2005}. In the highest luminosity state, the
source could have been observed in the \rosat data of 1991
November/1992 January, 1994 June and 1995 May, and certainly in the
\xmm observations of 2005 May and November. The source has thus been
detected in two epochs spaced by $\sim$8 years.  \cite{Kong2003} performed
a deep analysis of the source and found it to be very luminous
($10^{38}$--$10^{39}\,\text{erg}\,\text{s}^{-1}$) and very soft
($kT\sim60$\,eV).  Using the most recent calibration files, we
re-evaluated the bolometric luminosity of SSS$_{1}$ to
$6.2^{+1.3}_{-3.9}\times10^{38}\,\text{erg}\,\text{s}^{-1}$ in
revolution 192 and
$3.3^{+0.7}_{-2.0}\times10^{38}\,\text{erg}\,\text{s}^{-1}$ in
revolution 195, at a confidence level of 99\%.

\cite{Read2001} reported that the count rate of SSS$_{1}$ in 1992 May
($7.4\pm0.8\,\text{cts}\,\text{ks}^{-1}$) is equivalent to that of
1992 June ($7.5\pm0.8\,\text{cts}\,\text{ks}^{-1}$). This result
suggests that the duration of the outburst decline is several months
and that the decrease in luminosity within the 6\,days separating the
two first \xmm observations is just a short term flux modulation.  In
the light curve of SSS$_{1}$ during revolution 195, there are two
luminosity decreases lasting $\sim$5\,ksec, separated by $\sim$20\,ksec.
Using a Lomb-Scargle periodogram analysis \citep{Lomb1976,Scargle1982},
\cite{Kong2003} claim that the modulation present in the light curve
is periodic at a confidence level $>$99.9\% and conclude that this
periodicity could be associated with the orbital period of the system.
Assuming white noise variability, however, is not an adequate
assumption for X-ray binaries where many systems show strong
variability on timescales of hours.  Using the Monte Carlo approach of
\cite{benlloch:01a}, we re-evaluated the confidence level assuming a
red-noise process instead of pure white noise. We found that the
5.4\,h period is significant only at the 68\% confidence level.  Much
longer observations than the existing ones are therefore required
to be able to associate the 5.4\,h feature with some periodic
signal, which might or might not be related to the orbital period of
the system.

The properties of SSS$_{1}$ make it a very similar source to
SSS$_{2}$.  They both can be modelled with absorbed blackbodies with a
temperature of $\sim60\,$eV, they both present transient behaviour,
and their maximal bolometric luminosities observed in X-rays
are in both cases close to $10^{39}\,\text{erg}\,\text{s}^{-1}$.  The
SSSs in NGC~300 therefore have a luminosity in their high state that
classifies them as intermediate between the well-known `classical'
SSSs and the ultraluminous SSSs. As we discussed in
Sect.~\ref{sec:int}, this kind of system has not been well studied and
only a few such sources have been reported.  \cite{Orio2005} 
observed a variable SSS in M~31, r3-8, with a luminosity in
the high state at $\sim6\times10^{38}$\,erg\,s$^{-1}$.  Two other SSSs
were observed in M~81 and one in M~101
\citep{Swartz2002,DiStefano2003}, all with
$L_\text{bol}\sim4\times10^{38}\,\text{erg}\,\text{s}^{-1}$.  The few
observations of these SSSs, however, cannot establish a possibly
transient behaviour nor the highest luminosity levels reached by the
sources.

\subsection{Interpretations for the high and soft state of the sources}
The most natural explanation for these SSSs would be a steady nuclear
burning of hydrogen accreted onto a white dwarf (WD). However, as
shown in the previous section, the bolometric luminosity is above the
Eddington limit for a $1.4\,\text{M}_\odot$ compact object
($1.82\times10^{38}\,\text{erg}\,\text{s}^{-1}$). To explain the
nature of the variable ultraluminous supersoft X-ray source in the
Antennae, \citet{Fabbiano2003} suggested beamed
emission from nuclear burning onto a WD, with a beaming factor,
$b=L/L_{\text{sph}} = 0.01$ , where $L_{\text{sph}}$ is the inferred
isotropic luminosity of the blackbody and $L$ the true source
luminosity.  They suggest that the most likely cause of the anisotropy
would be a warping of the accretion disk.  In our case the beaming
factor would be only of $\sim$0.25.

Another model, suggested by \citet{Kong2003}, \citet{Kong2004}, and
\citet{Swartz2002} to explain the luminous SSS(s) present in NGC~300,
M~101 and M~81, respectively, is the presence of an intermediate-mass
black hole in the source. Assuming a blackbody model, with
$kT\sim60$\,eV, a luminosity of
$1\times10^{39}\,\text{erg}\,\text{s}^{-1}$, and assuming that the
X-ray emission comes from the innermost stable orbit, \citet{Kong2003}
estimate the mass of the black hole to $\sim 2800\,\text{M}_{\odot}$.
For our SSS, this hypothesis seems very unlikely: when observed in the
`high' state, this massive black hole would emit only at $<0.3\% 
$ of its Eddington limit ($3.64\times10^{41}\,\text{erg}\,\text{s}^{-1}$),
and much lower in the quiescent state. However, as reported by
\cite{Nowak1995} for stellar-mass black holes, below a few percent of
the Eddington luminosity, the sources are dominated by hard
non-thermal emission and soft emission is only observed once the
source luminosity increases by several percent of the Eddington
luminosity.

Another model was suggested by \cite{Mukai2003} and
\cite{Fabbiano2003} to explain the SSS ULX in M~101 and in the
Antennae, respectively. In this model, when material is accreted above
the Eddington rate, the excess of matter is ejected from the inner
part of the disk. The electron scattering opacity induced by the
wind/outflow involves supersoft blackbody emission from a photosphere
of $10^8$--$10^9$\,cm. \cite{King2003} re-analysed this model in more
detail: assuming a radial outflow with an outflow rate
$\dot{M}_{\text{out}}$ in a double cone occupying a solid angle $4 \pi
b$, at a constant speed, they showed that the outflow is Compton-thick
for $\dot{M}_\text{out}\sim\dot{M}_\text{Edd}$. This result is true
for $b\sim1$, when scattering of photons from the sides of the outflow
is negligible, and for $b\ll1$, when scattering is dominant.  The
emission is therefore mainly thermalized and observed as a soft
spectral component \citep{King2003}. The authors also evaluated the
temperature of this soft blackbody component as:
\begin{equation}
T_\text{eff}=1\times10^5 g^{-1/4} \dot{M}_1^{-1} M_8^{3/4}\,\text{K}
\label{equ:teff}
\end{equation}
where $g(b)=1/b$ or $1/(2b^{1/2})$ (for $b\sim1$ or $b\ll1$),
$\dot{M}_1=\dot{M}_{\text{out}}/(1\,\text{M}_{\odot}\text{yr}^{-1})$,
and $M_8=M/10^8\,\text{M}_\odot$, and $M$ is the mass of the
  accretor.  
These results confirm the hypothesis and observations of
the SSS ULX from \cite{Mukai2003} and \cite{Fabbiano2003}.
For the SSSs in NGC~300, the accretion rate $\dot{M}=L/(\eta
  \text{c}^2)$, where $L$ is the luminosity and $\eta$ the radiative
  efficiency, is $\sim1.4\times\,10^{-7}\text{M}_\odot
  \text{yr}^{-1}$, if we take a typical value of $\eta=0.1$. This
  accretion rate value is not extreme considering that the system is
  observed in a high/outburst state.  Furthermore, assuming the source
  is close to the Eddington limit, where
  $\dot{M}_{\text{out}}\sim\dot{M}$, we are able to estimate the mass
  of the accreting object, using equation\,\ref{equ:teff}.  Fixing the
  temperature at 60~eV and the luminosity at
  8$\times$10$^{38}$\,erg\,s$^{-1}$, the mass is
\begin{equation}
  M=0.037 \left[ \frac{g^{1/4}}{\eta} \right] ^{4/3} \text{M}_\odot
\end{equation}
Except for very low values of $b$ ($\lesssim0.1$), $g^{1/4}$
  is between 1 and 2, while $\eta$ goes from 0.06 for a non-rotating
  black hole to 0.42 for an maximally rotating black hole. This
  results in a mass range between $\sim$$0.1\,M_\odot$ and
  $4\,M_\odot$, i.e., at the lower limit of the mass range for black
  holes in our Galaxy.

As a last model, we consider the viewing angle dependence of the
emission of supercritical accretion flows \citep{Watarai2005}, i.e.,
accretion flows with a mass accretion rate that is so large that the
accretion disk becomes geometrically thick due to enhanced radiation
pressure. In this case, if the binary system is viewed at high
inclination, the outer part of the disk occults its inner parts, such
that only the very soft spectrum from the outer disk is observable.
For such a flow around a $10\,\text{M}_\odot$ black hole,
\citet{Watarai2005} showed that the emitted spectrum resembles  a
2\,keV blackbody at low inclination angle ($i<40^\circ$) and looks
like a 0.6\,keV blackbody at $i>60^\circ$.  Consequently, when viewed
edge-on, such a system will appear very faint. This is also the
argument given by \cite{Narayan2005} to explain why none of their 20
studied black hole X-ray binaries present eclipses. In our case,
although with 60\,eV the observed temperature is still lower than in
the example given by \citet{Watarai2005} for a $10\,M_{\odot}$ black
hole, we believe that this model is consistent with our observations.
When the outburst state observed in SSS$_2$ begins to decline, an inner
part of the disk would be visible, explaining the harder blackbody
component in the spectrum of rev.\ 1092.

%_____________________________________________________________________________

\section{Conclusions}
\label{sec:conc}
We report the discovery of a new luminous supersoft
source in the 2005 May \xmm observation of NGC~300. The
previously known luminous supersoft source detected by \rosat and the
previous \xmm data was below detectability.  This latter source
already appeared highly variable in the \rosat observations.

No object in the SIMBAD catalogue is associated with the new SSS, and,
from the optical data, no counterpart brighter than $\sim$21.7\,mag
($\sim$24.5\,mag for the previous known SSS) has been found.

The X-ray spectrum is well described by an absorbed blackbody at
temperatures of $kT$$\sim$60\,eV. The bolometric luminosity, in the
highest observed state, is 
$8.1^{+1.4}_{-4.5}\times10^{38}\,\text{erg}\,\text{s}^{-1}$ and dropped to
$2.2^{+0.5}_{-1.4}\times10^{38}\,\text{erg}\,\text{s}^{-1}$ six months
later.

The SSSs in NGC~300 are brighter than ``classical'' SSSs, for which
steady nuclear burning of hydrogen accreted onto white dwarfs has been
suggested to explain their nature. They are too faint, however, to be
classified as ultraluminous sources. We summarized several possible
explanations for their nature. These involve beaming emission from a
WD, intermediate-mass black hole (IMBH) or a stellar-mass black hole with
matter outflow or observed at high inclination angle. Except the one
including IMBH, which seems unlikely in our case, all models are
consistent with our observations.

\begin{acknowledgements}
  This paper is based on observations with \textsl{XMM-Newton}, an ESA
  science mission with instruments and contributions directly financed
  by the ESA Member States and the USA (NASA), and on observations
  made with ESO Telescopes at the La Silla observatory and retrieved
  from the ESO archive.  We acknowledge partial support from DLR grant
  50OX0002.  This work was  supported by the BMBF through
  the DLR under the project 50OR0106, by the BMBF through DESY under
  the project 05AE2PDA/8, and by the Deutsche Forschungsgemeinschaft
  under the project SCHN~342/3-1.
\end{acknowledgements}

\end{document}